\documentclass[final,5p,times,twocolumn]{elsarticle}
\makeatletter
\def\ps@pprintTitle{%
 \let\@oddhead\@empty
 \let\@evenhead\@empty
 \def\@oddfoot{\hfill\date{}}%
 \let\@evenfoot\@oddfoot}
\makeatother

\usepackage{tikz} 
\usetikzlibrary{shapes,arrows,automata,positioning}
\usepackage{circuitikz}

\usepackage{booktabs}
\usepackage{hhline}

\usepackage{subcaption}

\usepackage{subfiles}

\usepackage{url}

\usepackage{color}

\usepackage{blindtext}
\usepackage{microtype}
\usepackage{textcomp}

\usepackage[ngerman,english]{babel}
\usepackage{glossaries}
\usepackage[nolist,nohyperlinks]{acronym}

\usepackage{calc}
\usepackage{mathtools}
\usepackage{array}
\usepackage{amssymb}
\usepackage{amsthm}
\usepackage{amsmath}
\usepackage{upgreek} 

\usepackage{orcidlink} 

\newcommand{\unit}[1]{~\mathrm{#1}} 

\usepackage{scalerel}

\newacronym{sipm}{SiPM}{Silicon-Photomultiplier}
\newacronym{sipms}{SiPMs}{Silicon-Photomultipliers}
\newacronym{geant}{GEANT4}{Geometry And Tracking 4}
\newacronym{desy}{DESY}{Deutsches Elektronen-Synchrotron}
\newacronym{luxe}{LUXE}{Laser Und XFEL Experiment}
\newacronym{eds}{EDS}{Electron Detection System}
\newacronym{slac}{SLAC}{SLAC National Accelerator Laboratory}
\newacronym{facet}{FACET-II}{Facility for Advanced Accelerator Experimental Tests}
\newacronym{e320}{E-320}{E-320 experiment}
\newacronym{qed}{QED}{quantum electrodynamics}
\newacronym{sfqed}{strong-field QED}{strong-field quantum electrodynamics}

\let\as\acrshort
\let\al\acrlong
\let\af\acrfull

\definecolor{ngray}{gray}{0.6}
\definecolor{ngreen}{RGB}{141,174,16}
\definecolor{nblue}{RGB}{0,166,235}

\begin{document}

\begin{frontmatter}

\title{A High-Flux Electron Detector System to Measure\\Non-linear Compton Scattering at LUXE}


\author[DESY,UHH]{Antonios Athanassiadis\orcidlink{0009-0008-9963-0024}\corref{correspondingauthor}}
\cortext[correspondingauthor]{antonios.athanassiadis@desy.de}

\author[SLAC]{Robert Ariniello}

\author[DESY]{Louis Helary\orcidlink{0000-0001-7891-8354}}

\author[DESY,UCL]{Luke Hendriks}

\author[DESY]{Ruth Jacobs\orcidlink{0000-0001-5446-5901}}

\author[SLAC]{Alexander Knetsch}

\author[DESY]{Jenny List\orcidlink{0000-0002-0626-3093}}

\author[EPT]{Sebastian Meuren\orcidlink{0000-0002-2744-7756}}

\author[UHH]{Gudrid Moortgat-Pick}

\author[SLAC]{Ivan Rajkovic}

\author[DESY]{Evan Ranken}

\author[PULSE]{David A. Reis}

\author[DESY]{Stefan Schmitt}

\author[DESY]{Ivo Schulthess\orcidlink{0000-0002-5621-2462}}

\author[SLAC]{Doug Storey}

\author[PULSE,UNE]{Junzhi Wang}

\author[DESY,UCL]{Matthew Wing}

\address[DESY]{Deutsches Elektronen-Synchrotron DESY, 22603 Hamburg, Germany}
\address[EPT]{LULI, CNRS, CEA, Sorbonne Université, Ecole Polytechnique, Institut Polytechnique, 91128 Palaiseau, France}
\address[SLAC]{SLAC National Accelerator Laboratory, Menlo Park (CA) 94025-7015, USA}
\address[PULSE]{Stanford PULSE Institute, SLAC National Accelerator Laboratory, Menlo Park (CA), USA}
\address[UHH]{Universität Hamburg, 20148 Hamburg, Germany}
\address[UCL]{University College London, London WC1E 6BT, United Kingdom}
\address[UNE]{University of Nebraska–Lincoln, Lincoln (NE) 68588, USA}

\begin{abstract}
Recently, advancements in high-intensity laser technology have enabled the exploration of non-perturbative Quantum Electrodynamics (QED) in strong-field regimes. Notable aspects include non-linear Compton scattering and Breit-Wheeler pair production, observable when colliding high-intensity laser pulses and relativistic electron beams. The LUXE experiment at DESY and the \al{e320} at SLAC aim to study these phenomena by measuring the created high-flux Compton electrons and photons. We propose a novel detector system featuring a segmented gas-filled Cherenkov detector with a scintillator screen and camera setup, designed to efficiently detect high-flux Compton electrons. Preliminary results from \as{e320} measurement campaigns demonstrate methods for reconstructing electron energy spectra, aiming to reveal crucial features of non-perturbative QED.
\end{abstract}


\end{frontmatter}


\section{Introduction}
In recent years, advances in high-power laser systems have opened new possibilities to investigate the almost uncharted regime of \af{sfqed}. These strong fields occur in astrophysical theories of magnetars, within crystal lattices~\cite{nielsen_differential_2023} or in future high-energy particle colliders. In the presence of strong fields \as{qed} becomes non-perturbative, where the virtual electron-positron pairs become real, which leads to many non-linear, rarely measured phenomena. In order to enter the strong-field regime an electrical field, $\epsilon$, larger than the Schwinger critical field, 
\begin{align}
    \epsilon = \frac{m_e^2c^3}{e\hbar} = 1.32~\times~10^{18}\unit{V/m}
    \label{eq:schwinger_fieldstrength}
\end{align}
\noindent is needed~\cite{fedotov_advances_2023}. As modern technologies cannot achieve such an intense field strength, relativistic probes can be used in combination with a high-intensity laser. Because the electrical field strength is not Lorentz invariant, a boosted probe experiences larger fields within its laboratory frame. In the running \af{e320}~\cite{reis_e-320_2024} or planned \af{luxe}~\cite{abramowicz_conceptual_2021, luxe_collaboration_technical_2024} high-energy electrons ($\mathcal{O}~(10\unit{GeV})$) are being used in combination with a high-power laser system ($\mathcal{O}~(10-100\unit{TW})$) to probe \as{sfqed} theories. Due to the low interaction probability, high-intensity beams ($\mathcal{O}~(10^{9}\unit{/~Bunch~Crossing})$) are needed, creating the demand for particle detector systems with large dynamic ranges. 
\begin{figure}[t!]
    \centering
    \includegraphics[width=0.9\linewidth, trim={15mm 55mm 15mm 55mm}, clip=true]{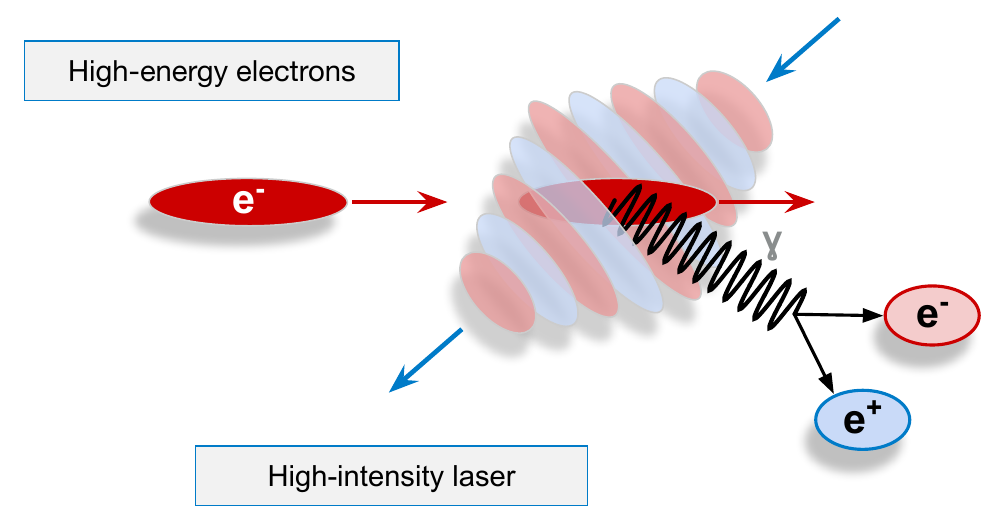}
    \caption{Schematics of an electron bunch from the European XFEL interacting with a high-intensity laser pulse. A high-energy photon is created via the non-linear Compton effect and interacts further with the field to create a Breit-Wheeler electron-positron-pair.}
    \label{fig:e-laser}
\end{figure}

\section{The Laser Und XFEL Experiment}
The \as{luxe} Experiment is a planned experiment at DESY and the European XFEL using its $16.5\unit{GeV}$, $250\unit{pC}$ electron bunches for interactions with a laser pulse peak power of $40$ to $350\unit{TW}$ (Fig.~\ref{fig:e-laser}). Two out of many processes that can be observed after the collision are the non-linear Compton scattering and the Breit-Wheeler pair creation. In the presence of the large electrical field, an electron interacts with multiple laser photons, resulting in a Compton electron and a high-energy photon. This photon can then furthermore undergo a Breit-Wheeler process within the laser field, creating an electron-positron pair. \\

\begin{figure}[t!]
    \centering
    \includegraphics[width=0.9\linewidth, trim={5mm 5mm 5mm 5mm}, clip=true]{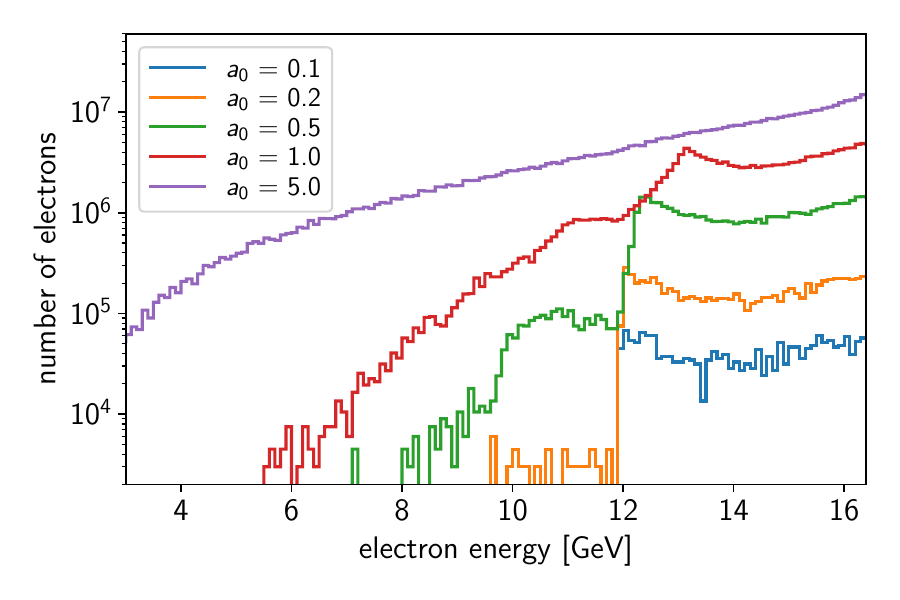}
    \caption{\textit{Ptarmigan} simulations of electron-laser collisions for different laser intensities $a_0$. The number of scattered Compton electrons is given per bunch crossing with a bin size of $100\unit{MeV}$. The blue line ($a_0 = 0.1$) corresponds to the case of linear Compton scattering.}
    \label{fig:spectrum}
\end{figure}

The \textit{Ptarmigan}~\cite{blackburn_simulations_2023} simulations presented in Fig.~\ref{fig:spectrum} show Compton electron energies per bunch crossing resulting from electron-laser collisions. For increasing field strengths, therefore increasing laser intensities $a_0$, a shift of the Compton edge towards higher energies and multiple harmonics can be observed. \as{luxe} will measure these effects with high statistics and accuracy such that non-perturbative theories can be verified experimentally. 

\section{The Electron Detection System}
\begin{figure}[h!]
    \centering
    \includegraphics[width=0.9\linewidth]{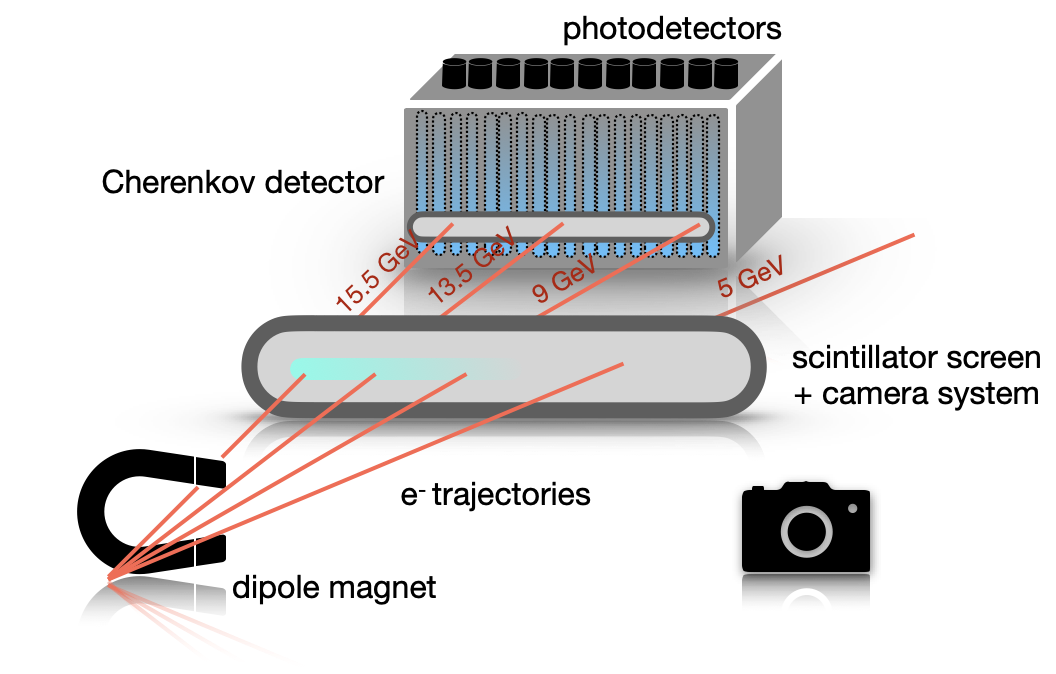}
    \caption{Sketch of the \al{eds} at \as{luxe} behind a spectrometer dipole magnet. It shows the two complementary systems: the scintillating screen and camera system as well as the Cherenkov-based straw detector~\cite{luxe_collaboration_technical_2024}.}
    \label{fig:eds}
\end{figure}

In order to measure the particle flux and the energy distributions of the particles in these interactions, a variety of detection methods are needed. This work focuses on measuring the electron energy spectra of the scattered electrons from the non-linear Compton process. The \af{eds} serves, in combination with a dipole magnet, as an electron energy spectrometer. Following the interaction point, a spectrometer dipole magnet separates the electrons from other particles with respect to their energy, allowing energy measurements with a spatially segmented detector system. As shown in Fig.~\ref{fig:eds}, \as{eds} consists of two complementary detector systems that use two different principles (scintillation and Cherenkov effect) for cross-calibrations and minimal uncertainties. \\

The scintillating screen and camera system (here: 'screen detector') consists of a terbium-doped gadolinium-oxysulfide scintillating screen with an active area of $100\times100\unit{mm^2}$ and a thickness of $0.05\unit{mm}$ as well as a \textit{Basler} camera ~\cite{basler_ag_basler_2025} which is commonly used in such detector configurations. When electrons travel through the screen, visible light is emitted via the scintillation process. The intensity is proportional to the number of electrons passing through. Because electrons are separated spatially by energy, an intensity measurement along the screen results in a measurement of the electron energy distribution. The screen detector will have a spatial resolution of $< 0.5\unit{mm}$ which results in a relative energy resolution of about~$2\%$. \\

\begin{figure}[h!]
    \centering
    \includegraphics[width=0.55\linewidth]{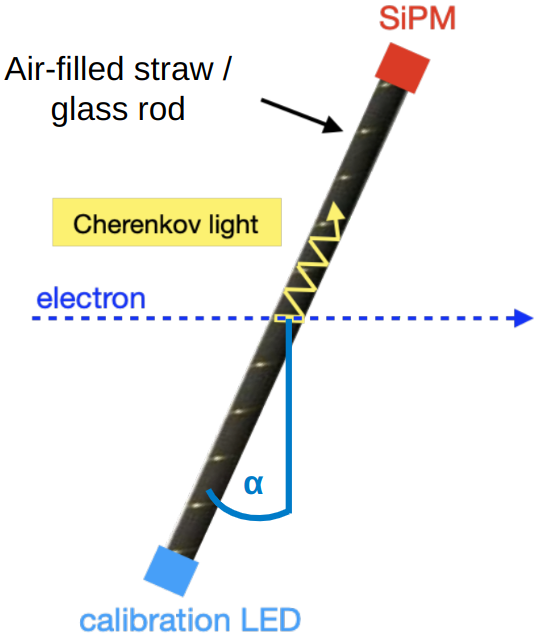}
    \caption{Sketch of a straw in the straw detector. The Cherenkov light (\textit{yellow}) created by the electron (\textit{blue arrow}) is reflected towards the top of the straw and detected by a \as{sipm}. Tilting the straws by an angle $\alpha$ can be used to tune the sensitivity.~\cite{luxe_collaboration_technical_2024}.}
    \label{fig:straw}
\end{figure}

The Cherenkov-based straw detector (here: 'straw detector') contains hollow stainless-steel tubes and solid quartz glass rods ('straws'), which are mechanically connected to a \af{sipm}~\cite{hamamatsu_photonics_kk_hamamatsu_2023} on the one side and a LED pulser system for calibrations on the other side, as indicated in Fig.~\ref{fig:straw}. 

When electrons pass through the straw volume, they produce Cherenkov light in the medium that is reflected at the inside walls of the straws towards a \as{sipm} sitting on top. The number of Cherenkov photons $N$ with frequency $f$ generated by a charged particle with velocity $v$ and charge $z$ in a medium of length $x$ is described in eq.~\eqref{eq:ncherenkov}. The Cherenkov angle $\Theta_\mathrm{C}$ is proportional to the refractive index of the material $n$ and the velocity of the charged particle $v$~\cite{leroy_principles_2016}. \\

The Cherenkov medium affects the number of optical photons produced by electrons passing through the straw, since the refractive index of air is $n_{\mathrm{air}}\approx~1.00027$) compared to that of glass $n_{\mathrm{glass}}\approx~1.4$. 
\begin{align}
    \frac{dN}{dx} &= \int_{f_1}^{f_2}\frac{2\pi\alpha z^2}{c}~\sin^2(\Theta_\mathrm{C})~\mathrm{d}f; ~~~~\Theta_\mathrm{C} = \frac{1}{\beta n};~~~~\beta = \frac{v}{c}\label{eq:ncherenkov}
\end{align}
Since the number of electrons passing each straw is expected to vary (see spectral distribution in Fig.~\ref{fig:spectrum}) a change of medium depending on the expected particle rate can extend the dynamic range of the detector. In this study, two different media have been tested: air and quartz glass. Furthermore, tilting the straws at an angle $\alpha$ (Fig.~\ref{fig:straw}) affects the absorption of the Cherenkov light on its trajectory towards the \as{sipm}.

\section{The Data Taking Campaign at FACET-II}
\begin{figure}[b!]
    \centering
    \includegraphics[width=0.9\linewidth, trim={8mm 40mm 3mm 40mm}, clip=true]{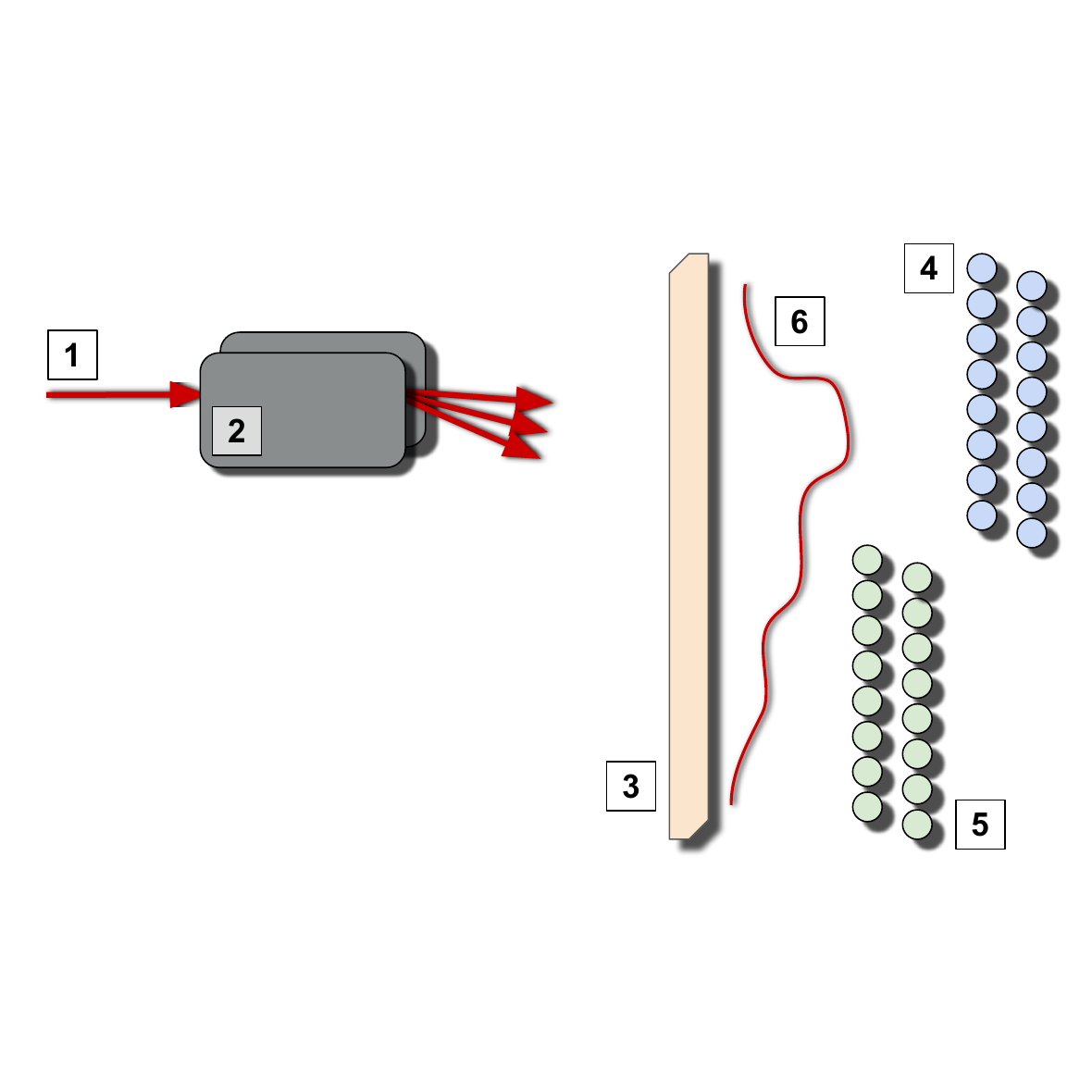}
    \caption{Schematics of the \as{eds} prototype setup at the \as{facet} beamline. The $10\unit{GeV}$ electrons and scattered electrons from electron-laser interactions~(1) are separated spatially with respect to their energy via a spectrometer dipole magnet~(2), bent downwards and then detected by the \as{eds} prototype in the scintillating screen~(3), stainless-steel straws~(4) and glass rods~(5). As sketched in~(6), a Compton spectrum is expected to be seen along the vertical axis.}
    \label{fig:luxe320sketch}
\end{figure}

For the purpose of testing the existing detector design, an \as{eds} prototype has been developed such that it suits the experimental program of the \al{e320} at the \as{facet} facility at the \al{slac} in Menlo~Park/CA, USA. The electron bunches ($10\unit{GeV}$, $1.6\unit{nC}$) collide with a $10\unit{TW}$-laser, creating an $a_0\lesssim 5$. Behind the interaction point, three imaging quadrupole magnets and a spectrometer dipole magnet transport the electrons about~$25\unit{m}$ downstream the accelerator to an in-air detector area right before the main beam dump. $16$ air-filled stainless-steel straws and $16$ solid glass rods are placed horizontally as shown in Fig.~\ref{fig:luxe320sketch}. \\ 

\begin{figure}[t!]
    \centering
    \includegraphics[width=0.9\linewidth, trim={5mm 5mm 5mm 5mm}, clip=true]{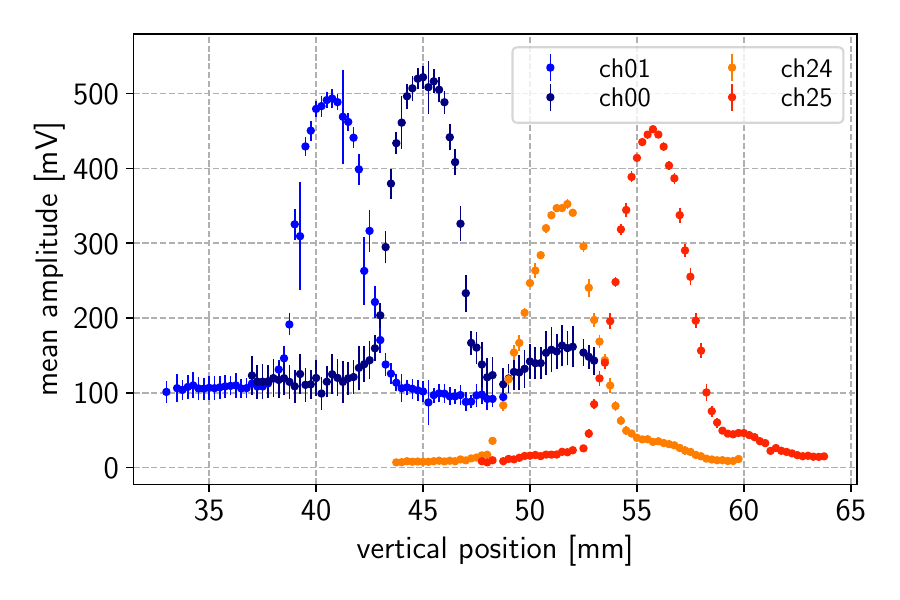}
    \caption{Mean \as{sipm} signal amplitudes of two air-filled steel straws~(ch$00$~\&~ch$01$) and two glass rods~(ch$24$~\&~ch$25$) as a function of the vertical detector position in the mono-energetic beam.}
    \label{fig:yscanplot}
    \vspace{-0.4cm}
\end{figure}

In the experimental campaign, the signal of the straw detector was calibrated. For this, the detector was moved through a Gaussian-shaped pencil beam along the vertical axis. Figure~\ref{fig:yscanplot} shows the change of signal amplitude in some of the straws with respect to the position of the detector in the main beam. The \as{sipms} attached to the air-filled straws were operated at a gain of about $3\times10^5$ and the \as{sipms} of the glass rods at a gain of approximately $1.5\times10^5$.


\newpage

\begin{figure}[h!]
    \centering
    \includegraphics[width=0.9\linewidth, trim={5mm 5mm 5mm 5mm}, clip=true]{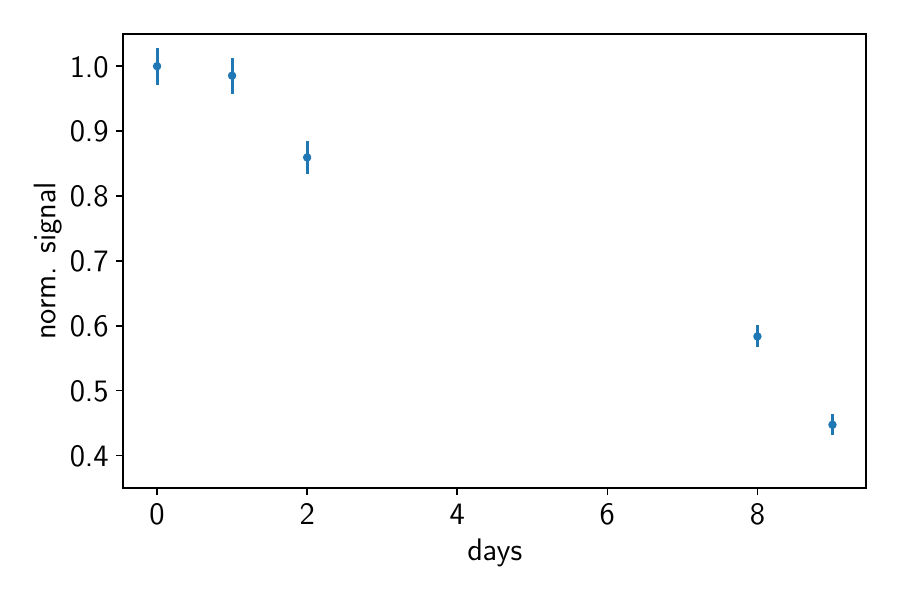}
    \caption{The normalized signal amplitude from LED pulser measurements over time which indicates a degradation of the detector due to radiation.}
    \label{fig:raddamplot}
\end{figure}

\noindent Three statements can be made from the measurement shown in Fig.~\ref{fig:yscanplot}:
\begin{itemize}
    \item By setting different \as{sipm} gains, the $1.6\unit{nC}$ charge can be successfully measured in both the air-filled straws and the glass rods.
    \item The Gaussian beam width can be measured to $(1.3~\pm~0.2)\unit{mm}$, which is in agreement with the machine setup.
    \item When not aiming at one straw, a clear signal can be measured, which results in the vertical offset of the Gaussian-like graphs. This background originates from signals created by charged particles hitting the sensitive area of the \as{sipms} directly.
\end{itemize}
\vspace{0.3cm}
Throughout the initial measurement campaign at \as{facet}, severe radiation damage was observed in the \as{sipms}. With the constant LED light-yield as reference, a reduction in the \as{sipm} signal amplitude is visible, which is a typical sign for radiation damage in \as{sipms}~\cite{garutti_radiation_2019}. Within approximately $9\unit{days}$ of accelerator operation the signal degraded by about $50\unit{\%}$ (Fig.~\ref{fig:raddamplot}).

\begin{figure}[h!]
    \centering
    \includegraphics[width=0.9\linewidth, trim={5mm 5mm 5mm 5mm}, clip=true]{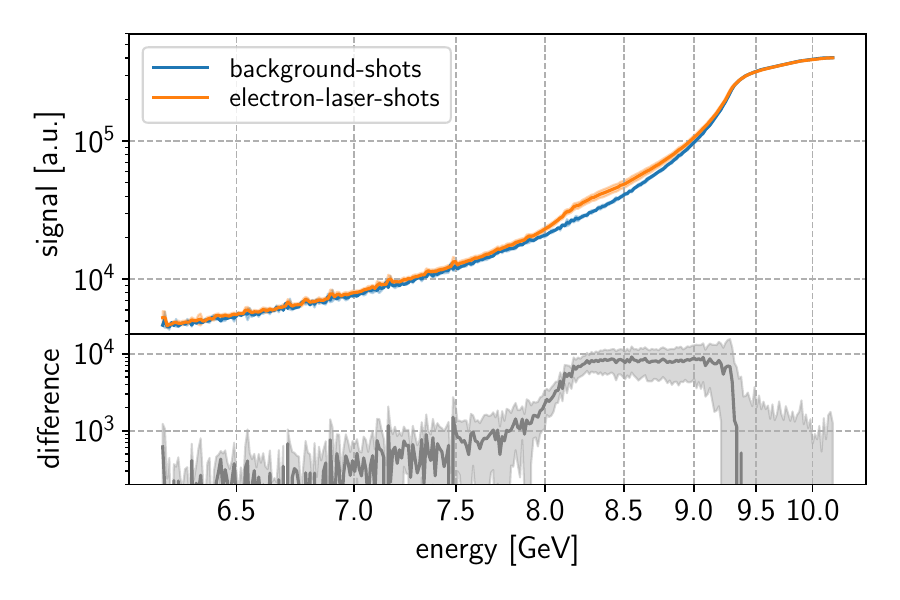}
    \caption{Compton electron energy spectrum of electron-laser collisions in the \al{e320} recorded with the screen detector. The signal~(orange) and the background~(blue) are averages over $\mathcal{O}~(10)$ shots with a $1\sigma$ band. The background signal is generated every $10\unit{bunches}$, where no electron-laser collisions take place~\cite{reis_e-320_2024}. In gray the difference between the signal and the background is shown.}
    \label{fig:screenspectrumplot}
\end{figure}

\newpage
When observing electron-laser collisions in the \al{e320} the scintillating screen shows a clear Compton electron energy spectrum (Fig.~\ref{fig:screenspectrumplot}). In comparison to the background, where no laser collisions occurred (hence, only the main beam passes through the screen), a Compton edge becomes visible around roughly $8\unit{GeV}$.

\section{Conclusion \& Outlook}
The \al{eds} is a detector that functions as an electron energy spectrometer with high particle flux compatibility. It consists of two complementary systems, a scintillating screen with a camera system and a segmented Cherenkov-based straw detector. First measurements at the \as{facet} facility and in cooperation with the \as{sfqed} experiment \as{e320} gave new insights into the performance of the detector to measure a Compton spectrum in a running \as{sfqed} experiment and the radiation effects in a real experimental environment.
\\
In the future, further data analysis, i.e. combining straw and screen data, will provide a deeper insight into the described phenomena and detailed \as{sfqed} and Monte Carlo simulations allow for a comparison of the measured electron spectra. Optimizations like a more radiation-hard design or stricter background suppression will bring the \al{eds} one step closer to a final detector system for the \as{luxe} experiment at the European XFEL.

\section*{Acknowledgment}
I gratefully acknowledge the \as{e320} collaboration and the \as{facet} team, whose on-site support at \as{slac} made our measurement campaign possible. For technical support at \as{desy}, I thank K. Gadow and M. Klotz. 
The FACET-II accelerator test facility at SLAC is supported by the U.S. Department of Energy contract DE-AC02-76SF00515. 
D.~A.~Reis and J.~Wang are supported by the U.S. Department of Energy Office of Science, Office of Fusion Energy Sciences under award DE-SC0020076. S.~M.~Meuren is supported by the Centre national de la recherche scientifique (CNRS) and the Agence Nationale de la Recherche (ANR) under the Chaire de Professeur Junior (CPJ) program. Furthermore, he is grateful for support from the visitor program of the Stanford PULSE institute.
This work was supported by the Quantum Universe Cluster of Excellence.

\section*{Statement}
During the preparation of this work the author used the Writeful language model in order to improve the readability and language of the manuscript. After using this tool, the author reviewed and edited the content as needed and takes full responsibility for the content of the publication.

\bibliography{references.bib}

\end{document}